# DFT Investigations of Major Defects in Quartz Crystal: Implications for Luminescence and ESR Dosimetry and Dating

Jalaja Pandya[1], Malika Singhal[2], Navinder Singh[1], Naveen Chauhan[2]

[1] Theoretical Physics Division, Physical Research Laboratory, Ahmedabad, India, 380 009.
[2] Atomic, Molecular and Optical Physics Division, Physical Research Laboratory, Ahmedabad, India, 380 009.

## Abstract

Quartz is extensively used for luminescence and ESR dosimetry as well as dating. These techniques use inherent defects introduced in quartz crystal during its crystallization in nature. The defect comprises both intrinsic as well as extrinsic defects. These defects give important luminescence properties to quartz but are not yet well understood from a theoretical perspective. Specifically, in case of luminescence dosimetry the nature of traps and their involvement in luminescence production is not exactly known. Luminescence mechanism is broadly understood through some experimental observations which helped in deducing some models for explaining the luminescence mechanism. Thus, present work attempts to understand the basic physics of defects and their implication for luminescence and ESR techniques via Density Functional Theory (DFT) modelling. The work uses DFT to model the presence of some possible major impurities in quartz. Several interesting novel results are obtained that will have implications for ongoing research in Luminescence and ESR methods. The DFT modelling suggested that Oxygen deficiency in quartz crystal results in the formation of both electron and hole trapping centres. However, it is observed that these centres can be passivated by the introduction of charge compensating OH or H ions. Further, it is found that peroxy defects can be formed in the presence of either excess Oxygen or due to the absence of Silicon ($Si^{4+}$), however, the nature of the traps formed in both cases is different. Besides these intrinsic defects, Al and Fe are the major impurities which are observed as defects in quartz. The modelling of these impurities suggested that negligible change in DOS is observed for Al defect and Fe generally forms a recombination centre or hole trap. In addition to these, there are several interesting first-time observations that are not reported and will be helpful for progressing luminescence and ESR dosimetry research.

## I. INTRODUCTION TO QUARTZ AND LUMINESCENCE

Quartz ($SiO_2$) is a ubiquitous natural mineral which forms ~13% of the weight of the Earth [1], distributed everywhere on earth crust and useful for a number of applications in different industries. It is a versatile material useful for various domains of science and especially for the present work related to dosimetry and dating, the prime focus is related to luminescence (thermal/optical) studies of quartz.

Luminescence is the phenomenon of emission of light from an insulator followed by prior absorption of energy from high-energy ionizing radiations [2]. It is the property associated with defects of crystals, which results in the formation of allowed metastable states within the forbidden band of the crystal. These metastable states act as temporary resting sites for charge carriers (electrons and holes) and charge carriers can occupy these states for a certain amount of time, which depends on the lifetime of these states at a particular temperature. Depending upon the energy requirement and closeness of these centres to conduction band and Fermi-level decides the probability of eviction from these sites. They can be classified as traps or recombination centres (Fig-1). Charges are trapped in both trap centres and recombination centres however, charges trapped in trap centres have higher probability of eviction. Thus if such charges are stimulated by external energy sources like heat or light, they are evicted and become free to move in the crystal. The charges trapped in recombination centres are generally more stable as trapping sites are near the Fermi level and, far away from the conduction band. Their probability of eviction is low. These centres instead allow recombination of freely moving charge carriers with the opposite polarity charge carrier present (trapped) in recombination centres resulting in luminescence emission. The energy required for eviction of charge carriers (trap depth), lifetime, and other relevant crystal properties related to defect sites could be studied by understanding kinetics of luminescence process [3-7].

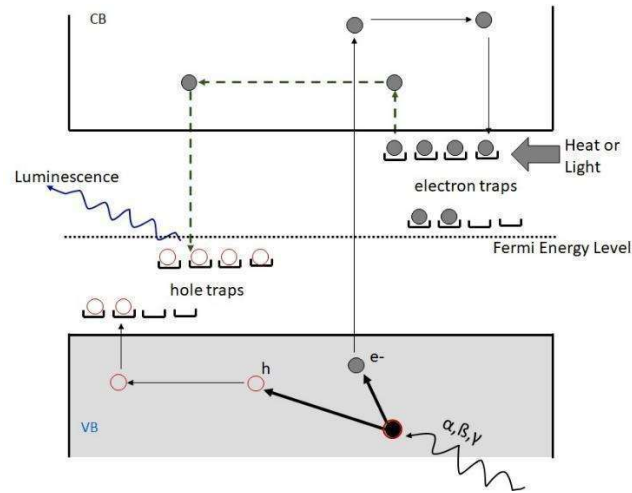

Fig. 1 Energy Band Model of Luminescence

Natural quartz exhibits interesting but quite variable luminescence properties. There are often several defect centres present in the quartz resulting in a range of trapping and recombination sites. These not only define the variable lifetimes but also controls emission spectrum of quartz luminescence. The lifetime of trapping sites in quartz is known to vary from a few hours to $10^8$ years [7] which are estimated based on the kinetic studies of luminescence in quartz. However, knowledge about the defects linked with luminescence in quartz is still unknown [8] and most of the predictions are based on limited or indirect correlation with electron paramagnetic resonance (EPR) spectroscopy, cathodoluminescence (CL) microscopy and spectroscopy, synchrotron X-ray absorption spectroscopy (XAS), -photoluminescence (PL) studies and others.

Natural quartz has a simple chemical formula but it is a complex system [9, 10]. It occurs in at least 15 different forms (polymorphs/crystal structure) depending upon P-T condition of crystallizations [1, 9]. Quartz is the most important silica polymorph in nature, and occurs as a common constituent of different types of magmatic, metamorphic and sedimentary rocks. Quartz primarily occurs in two forms: (1) the most prominent low temperature alpha quartz and (2) beta quartz that are stable at high temperature.

Quartz occurs as one of the purest minerals, however, trace impurities and defects get introduced during its crystallization due to ambient thermodynamic conditions and surrounding matrix environment. These impurities affect luminescence properties of quartz significantly [1, 2] and bear important information related to mineral genesis [11]. These impurities could be intrinsic or extrinsic point defects. The intrinsic defects could be primarily due to Si or O vacancies or displacements within lattice sites. The extrinsic defects could be due to presence of foreign ions in the lattice sites and interstitial positions. The point defects may further result in initiation of extended defects such as dislocation or planar defects [9]. $Si^{4+}$ has a small ionic radii (~0.42 Å) and it is rarely substituted by other elements due to its small size and high ionic charge. However, there are some ions like $Al^{3+}$, $Ga^{3+}$, $Fe^{3+}$, $Ge^{4+}$, $Ti^{4+}$ and $P^{5+}$, which can substitute $Si^{4+}$. *Among these, $Al^{3+}$ is the most common impurity and Fe being a major element in earth crust is another important impurity.* In order to compensate for the charge monovalent impurities like $H^+$, or alkali metal ($M^+$) like $Li^+$, $Na^+$ etc. are also often present [1] in the lattice. In case when charge is compensated by $H^+$, it forms OH bond, which can capture a hole to form a complex at low temperatures, and reduces to $[AlO_4]^0$ upon irradiation at room temperature [12] and imparts smoky colour to quartz [13, 14]. Similarly, Fe can be incorporated as $Fe^{2+}$, $Fe^{3+}$ and $Fe^{4+}$. As $Fe^{3+}$ has larger size, it distorts tetrahedral structure, so its incorporation is limited. $[FeO_4]^0$ complex formation imparts purple to pale colour to quartz and details concerning the $Fe^{4+}$ site and its local environment have not been determined very confidently [9].

Besides these, other trace element impurities are also observed. Impurities like $Ge^{4+}$ and $Ti^{3+}$ form ($Ge^{4+}/M^+$) and ($Ti^{4+}/H^+$) electron trapping centres [12]. In addition to these, there are defects, which are related to O vacancies or clusters. The group of oxygen-vacancy electron centres (OVEC) involves an oxygen vacancy (≡Si–Si≡) that  can be divided into the E' and E'' types, denoting one and two unpaired electrons , respectively [9]. There are E' centres (E'$_1$, E'$_2$, E'$_4$) [15] are most frequent defects other than Al centres. E'$_1$ centre is formed when an electron is trapped at Si site next to oxygen vacancy, E'$_2$ centre involves trapped proton associated with oxygen vacancy and electron at Si site.  Another centre E'$_4$ is an oxygen vacancy in which the hydride ion is bonded to Si atom and an electron is present at non-equivalent Si site [15].

The emission spectrum of these defect centres are often found to be quite broad and overlapping as a result it has been difficult to associate the emitted luminescence with their centres and the association established so far is only qualitative. The three main emission bands that are observed are 360-420 nm (near UV-violet), 460-480 nm (blue) and 610-630 nm (orange) [1]. The cathodoluminescence (CL) emission bands in blue and red/orange are considered to be due to the intrinsic defects [9].  However, different studies interpret the emissions differently and a conclusive identification is not yet established. The source of the UV-violet band is uncertain. Some works attributed it to recombination of electrons with holes trapped at $H_3O_4$ centres [16] while others relate this emission with electron-hole recombination at $[AlO_4]^0$  or at $[AlO_4/M^+]^+$ centre [12]. The blue emission is often linked with $[AlO_4]^0$ centre [13, 14, 16]. The broad TL emission band cantered at 560- 580 nm was observed only in natural quartz of hydrothermal origin [17].

Quartz thermoluminescence (TL) glow curve often have glow peaks at several temperatures [1, 5-7, 12, 13]. Most prominent among these are peaks corresponding to temperatures of 110°C, 240°C, 325°C and 370°C. Out of these, 325°C is considered optically bleachable and often used for optically stimulated luminescence (OSL) dating. The origin and linkages of these peaks is not well established and their association with the different centres is not yet well characterized. Kinetic studies of thermal/optically stimulated luminescence enable determination of trapping centre kinetic and dosimetry relevant parameters such as trap depth, frequency factor, saturation dose and minimum detectable dose, but nature of trap with which trapping and emission can be linked is not yet known. All these peaks are found to emit in a range of emission bands [1, 12, 18] out of which only few are considered stable over long times and thus used for studies related to dosimetry or chronology. This lack of certainty related to identification of the defects involved in the luminescence processes has led to corresponding uncertainty in describing the TL and OSL emission mechanisms. Further to this, the laboratory procedures are known to alter the stimulated luminescence intensity and the emitted spectra, the causes of which are yet not clear although several attempts are being made to understand it [5, 7]. However, this can be understood if the linkage of luminescence and respective centres can be established.

The discussion so far suggests that although quartz is a simple mineral, the defects associated with it are quite complex and they modulate the luminescence and other physical properties quite effectively. Thus, it is important to understand the effects of defects on quartz crystal structures and its energy levels, which can be linked to the physical, spectral and luminescence properties of the quartz. If this can be done via basic theoretical physics modelling, it can provide quite useful information to support experimental observations. However, exact theoretical computations are quite complex, but some preliminary computation can possibly suggest the future course of improvement. Here comes the advantage of studying defects in quartz using the density functional theory (DFT) which we take up in this study. *As an initiation of this aspect, present work tries to explore the potential of DFT calculations for understanding the effect of defects in quartz on its band structure and some physical quantities.* Therefore, in this work, study is limited to very basic computations considering only most probable defect centres that are expected to be present in quartz. Some of these could be the presence or absence of constituting atoms of the crystal lattice or substitution of these atoms by other major elements present in earth's crust during its crystallization. The work also tries to draw some crude linkages in luminescence-based observations and DFT based results, which may require refinement in future as scientific understanding of DFT simulations and defects in quartz is gained.

## II. Introduction to DFT

DFT is a quantum mechanical approach widely used to investigate electronic structure and other physical properties of molecules and condensed matter [19-21]. In this, the single electron ground state density $\rho(r)$ is used to calculate the electronic structure of the system instead of wavefunction $\phi$ which is used in conventional quantum mechanics [22]. In 1964, the work of Hohenberg and Kohn proved the exactness of this approach (using $\rho(r)$ as the fundamental variable) [23]. The ingenious work of Kohn and Sham [24] laid the theoretical foundation of DFT. Kohn and Sham [24] reformulated the interacting many-body problem to an equivalent single particle formalism [25, 26]. All the present day DFT models are based on this Kohn-Sham formalism. The only unknown functional is the Exchange-Correlation functional $E_{XC}[\rho]$ (equation 1).

$$E_{Kohn-Sham}[\rho] = T[\rho] + E_H[\rho] + E_{XC}[\rho] + \int \rho(\vec{r})V_{eff}(\vec{r})d^3r \qquad (1)$$

In the above equation, $T[\rho]$ is the kinetic energy functional, $E_{XC}[\rho]$ the Hartree functional and $V_{eff}(\vec{r})$ is the effective potential of the fictitious single particle system. This self-consistent Kohn-Sham equation is solved by using the set of orbitals called Kohn-Sham orbitals, which gives the electronic density of the interacting real system [27].

The exactness of DFT depends on the accuracy of the exchange correlation functional [24]. Since the explicit form of $E_{XC}[\rho]$ is not known, some sort of approximations are used for it [28]. The local density approximation (LDA) is the most simplest form of $E_{XC}[\rho]$, wherein $E_{XC}[\rho]$ is approximated only in terms of the electron density $\rho(r)$ [29]. The generalized gradient approximation (GGA) functions, which are more accurate, takes into account density gradient $\nabla\rho(r^\sim)$ dependence as well [30]. Inaccurate cohesive energies in molecules and solids obtained using LDA are considerably corrected in GGA. In the current study, the GGA of Perdew-Burke-Ernzerhof based $E_{XC}[\rho]$ is used for the DFT calculations.

Since the last few decades, DFT has become a practical tool in understanding the physical and chemical properties of systems at molecular level. These calculations have also played an important role in characterizing defects and to study their effect on the physical properties in many condensed matter systems [31]. Its significant accuracy and computational efficiency makes it more feasible than other ab-initio methods. In addition, the availability of many DFT based software packages makes it popular. In this work, The Quantum Espresso Package (http://www.quantum-espresso.org) is used to perform all the DFT simulations. It uses the plane-wave pseudopotential method to calculate the properties of condensed matter systems. The codes are constructed using the periodic boundary conditions (PBC). PWscf code in this package calculates the total energy of the system. The minimum energy structure is obtained using Hellman and Feynman forces and stresses [32]. Essentially, the ground state properties of any element in any kind of atomic arrangement can be calculated.

### III. DFT settings For Quartz

In this work, the effect of the most commonly found defects in quartz on its electronic structure is theoretically investigated. The DFT simulations for pure quartz crystal were first performed to understand its energy states and bond structure. The structure of pure quartz was obtained from the materials project (https://next-gen.materialsproject.org/). It has a trigonal structure and belongs to the space group P3(2)21 [33]. At first, the minimum energy structure of quartz was determined (Fig.2 (a)) using its primitive cell containing 3 SiO$_2$ molecules (9 atoms). The unit cell parameters were optimized at zero pressure. The ionic cores were described using optimized norm-conserving Vanderbilt (ONCV) pseudopotentials [34]. The wavefunction and charge density cut-off were set to 90 Ry and 400 Ry respectively for all the structures, which is appropriate mainly for covalently bonded insulator systems and reproduces the properties of pure quartz very well (band gap etc.). A grid of 5 X 5 X 5 k points was used to sample the Brillouin zone. In the optimized structure of pure quartz there are two slightly different Si-O bond lengths, two having shorter bond length of ~1.61 Å and other two having longer bond length (~1.62 Å) [35]. The O-Si-O angle is ~111$^0$. The density of states (DOS) for pure quartz is shown in Fig.2 (b). Although the energy difference between the valence band maxima and conduction band minima (tail states) is ~ 5.27 eV, but the effective reported value of the experimentally observed band gap for pure quartz is about 8 eV. It is controlled by states in the tail of the conduction band, which have weight appreciably greater than the states in the exact bottom of the conduction band. We thus apply a criterion based on experiments for defining the cut-off DOS level. Band gap in quartz from optical experiments is estimated to be ~ 8-9 eV [36]. Thus, we take the energy gap between the points with the density of state ~6.5 (3.2) eV$^{-1}$, which is ~ 8.1 eV (dotted line in Fig-2(b)). Additionally, the spatial electron density and electric potential was also studied to understand the changes in bond structure and charge density redistribution due to incorporation of different defect centres. To understand the bond type, the atomic densities are subtracted from the total (pseudo) charge density. The electron density plot for pure crystal at isosurface level 0.04 a.u. is displayed in Fig-2(c). The electron density isosurface is observed between the Si and O atoms, indicating a covalent bond. In addition, as expected the electron density nearer to the O atoms as O is more electronegative than Si.

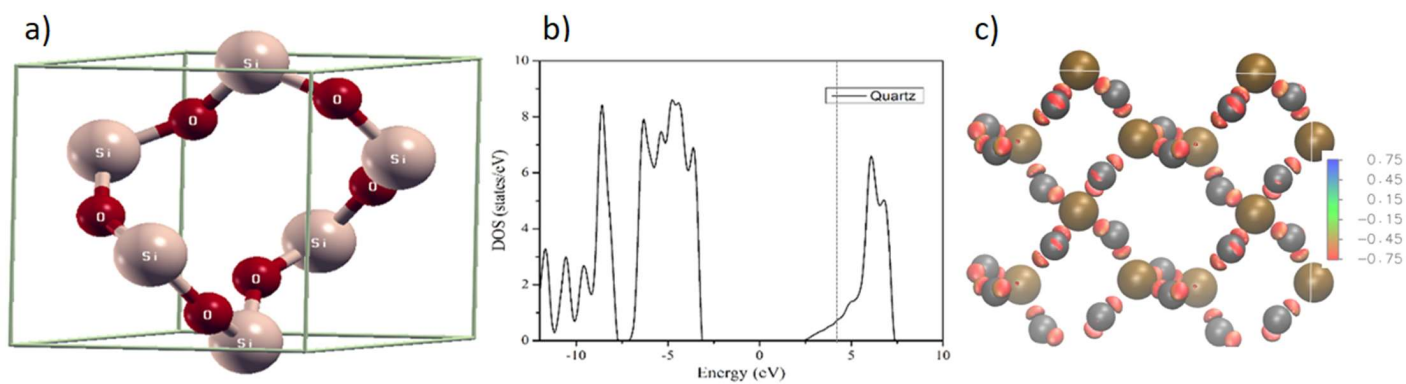

Fig-2 (a): The structure of a unit cell of quartz (SiO$_2$), where the small red balls represent the O atoms and large white balls are the Si atoms. (b) The density of state for the pure quartz system where 0 in the x-axis is set to the Fermi energy. (c) The electrostatic potential on the isodensity (0.04 a.u.) surface for the supercell of $SiO_2$. Red colour indicates negative potential.

The effect of defects in the electronic structure of quartz is addressed by carefully analysing the changes in the DOS of modified structure from pure crystal. In natural quartz present in sediment or rocks, defects are in trace amounts (ppm level), and defect-defect interactions can be neglected. Sufficiently large supercells of 1 X 2 X 2 unit cells containing 36 atoms are built and used for simulation for present study. The defects were later introduced in this supercell. Ideally simulating a 2 X 2 X 2 supercell instead of 1 X 2 X 2 is considered a better choice, but it takes a lot of computational time. In order to verify the effect on the result of choice of supercell, the DOS for oxygen vacancy defect in quartz was simulated using both the supercells and no qualitative difference was observed (Supplementary files) suggesting 1 X 2 X 2 will be sufficient for simulations.

The defect-containing supercell was optimized to obtain the minimum energy structure. All the optimization parameters were kept same. The defect structures were then optimized to zero pressure and DOS were then calculated for the optimized structures. Based on the change in DOS and their position in bandgap these defect structures are identified as electron/hole traps and recombination centres. Following section discusses the DFT computation for different major defects observed in quartz considering one defect at a time. The details of the major defects considered for simulations are provided in Table 1.

Table 1: Defects used for simulation using DFT in the quartz crystal

| Defect possibility | Possible defect combination | Previous reports | |
|---|---|---|---|
| Oxygen vacancy | Oxygen missing & Si-Si bond | $[O_3Si:SiO_3]$; Neutral O vacancy; [37] | This study |
| | Oxygen missing and an electron from Silicon missing | E' centre; [38] | This Study |
| | Oxygen missing and an electron from each of the two Silicon missing | None | This Study |
| | Oxygen missing & H attached to one silicon | $E'_2$ & $E'_4$ centre; [39] | This study |
| | Oxygen missing & an H attached to each of the two silicon | None | This study |
| | Oxygen missing & OH attached to Si | SiOH defect; [40] | This study |
| | Oxygen missing & 2OH attached to Si | SiOH defect; [40] | This study |
| Silicon vacancy | Silicon missing | None | This study |
| | Silicon missing: Oxygen forming peroxy bonds | None | This study |
| | Silicon missing: 4H replacing Si | SiOH defect; [40] | This study |

| | Silicon missing: 2H replacing Si and 1 peroxy bond | None | This study |
|---|---|---|---|
| | Silicon missing: 3H replacing Si and 1 1 oxygen left | None | This study |
| Substitutional defect | Silicon replaced by Aluminium +1H/Li/Na | $[AlO_4|H^+]^0$; [41] | This study |
| | Silicon replaced by Aluminium | $[AlO_4]^0$; [41] | This study |
| | Silicon replaced by Iron + 1H/Li/Na (Iron present in 3+ oxidation state) | $[FeO_4|H^+]^0$; [42] | This study |
| | Silicon replaced by Iron (Iron present in 3+ oxidation state) | $[FeO_4]^0$; [42] | This study |
| | Silicon replaced by Titanium | $[TiO_4]^0$; [41] | |
| | Silicon replaced by Titanium +H/Li/Na | $[TiO_4/H^+]^0$; [37] | |
| | Silicon replaced by germanium | $[GeO_4]^0$; [41] | |
| | Silicon replaced by germanium+H/Li/Na | $[GeO_4|Li^+]^0$; [43] | |
| | Silicon replaced by Phosphorous | $[PO_4]^0$; [37] | |
| | 2 Silicon replaced by Phosphorous and Aluminium | $[O_3POAlO_3]$; [44] | |
| Excess Oxygen | Peroxy linkage | $[O_3SiOO^-]$ | This study |

## IV. Defect Centres Studies: DFT Simulations, Observations and Interpretations

A discussion on the lattice structure modification, DFT computation setting, obtained results and inferences drawn in relevance to individual defects is given below.

### (A) Oxygen Vacancy Defect

O-vacancy is a commonly found defect in quartz. Experimentally, this defect is found to be stable in many conditions [45]. The neutral oxygen vacancy, which is also called Oxygen Deficiency Centre (ODC), is a diamagnetic defect and is formed when an O atom is missing leaving behind two electrons of neighbouring silica atoms. The ODC is believed to form both hole and electron trapping centres [46]. It is also a precursor of E′ and E′′ defects discussed in the introduction.

i) The optimized structure of quartz having such an O vacancy is displayed in Fig-3(a). The structure was obtained by removing one O-atom (marked by dotted square) from the supercell of quartz. It is observed that near ODC, the Si-Si bond length changes from 3.12 Å (pure quartz) to 2.41 Å, which is also reported by Zhang et al [47]. Thus, a weak bond is formed between two Si atoms to stabilize the defect. In addition, the Si-O bond lengths in the proximity of the defect slightly increase (from 1.61 Å to 1.65 Å). Fig-3(b) displays the DOS for O-vacancy crystal. A clear peak at 2.4 eV above the Fermi level of pure quartz near the conduction band is seen with 8.74 states/eV. This peak is 1.25 eV below our reference for conduction band bottom. The existence of the peak below conduction band in DOS indicates the presence of shallow electron trapping centres possibly with low lifetime. As the energy requirement for detrapping is small, the electrons in these traps can be excited by near infrared radiations (~1000 nm). The reported lifetime of a trap having trap depth ~1.42 eV with TL peak at 190 °C [7] is around 700 years suggesting the observed trap with 1.25 eV trap depth will have a lower lifetime. The presence of low-density band tail below CB states may also result in fading of the signal due to leaking of charges for overtime depending upon ambient temperature reducing the stability further. In addition to this, some new tail states are also observed near the valence band edge in the DOS.

ii) The spatial electron density and electric potential for this defect suggests that due to the presence of an O vacant site, the electron density in that region (marked by dotted square in Fig-3(a)) is considerably reduced. This can also be inferred from its density surface with isovalue 0.04 a.u., where only a tiny surface is seen between the Si-Si atoms. Further, the electric potential (plotted on the density isosurface) becomes slightly positive suggesting it to be a weak electron trap.

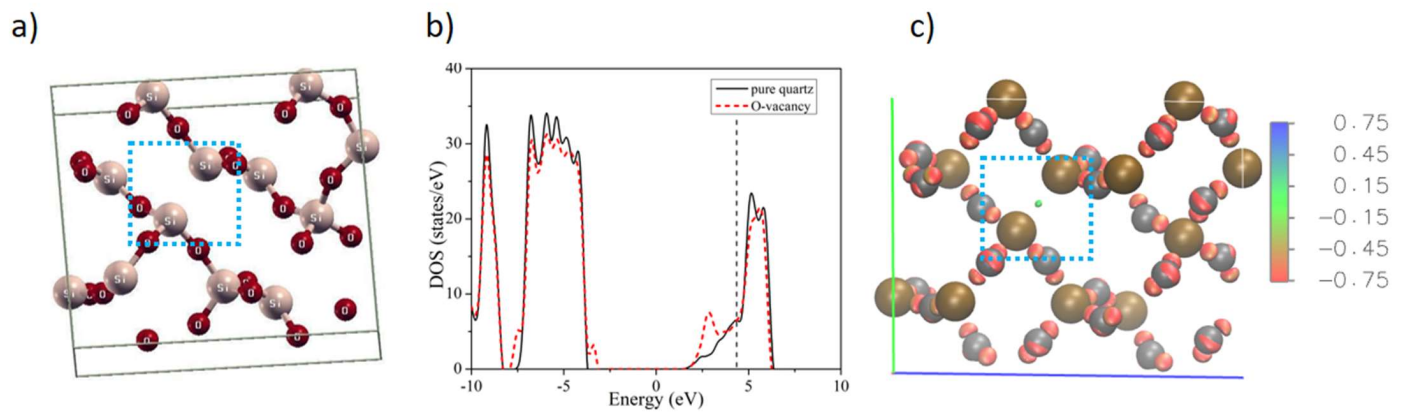

Fig-3 (a) The minimum energy structure of a 1x2x2 supercell with an oxygen deficiency centre. (b) The DOS plot for the defect. A peak at 2.75 eV above the Fermi level is observed. (c) The electrostatic potential on isodensity (0.04 a.u.) surface. The electron density near ODC reduces. Green colour represents positive electrostatic potential.

iii) Oxygen being an electronegative atom may also take an electron from a nearby Si atom. This type of defect is labelled as E' defect. This defect is simulated by giving an overall +1 charge to the supercell. E'$_1$ centre can be formed by various irradiations involving alpha particles, electron beams, neutrons, gamma- and X-rays [48] (Gotze et al 2021). Heating quartz till 573K leads to increase in concentration of E'1 by thermal activation. Above 573K, there is a loss of paramagnetic properties and E'$_1$ defect concentration decreases [49, 50]. This decay of the trap is due to trapping of electron [50].

iv) The minimum energy structure of this defect is displayed in Fig-4(a). It is quite similar to the ODC (Fig-3(a)). Here, the Si-Si length is 3.025 Å. The Si-O bond length near the defect is slightly reduced to 1.60/1.61 Å from 1.61/1.62 Å of pure quartz. In the DOS (Fig-4(b)) a new peak appears at ~0.3 eV below the reference. Two other small peaks are seen, one at ~1eV above the valence band maxima and other at ~2.6 eV below the reference level, both having ~5 states/eV DOS. The peak near VB is found to be matching with the reported activation energy (~0.95eV with lifetime $10^8$ years) [49]. However, there is no report about 2.6eV trap centre. Fig-4(c) shows the electron density plot with isovalue 0.008 a.u. with electrostatic potential plotted on top of it. Electron density ~0.008 a.u. is observed between the Si-Si atoms having positive potential. This is similar to that observed in neutral ODC but for low electron density.

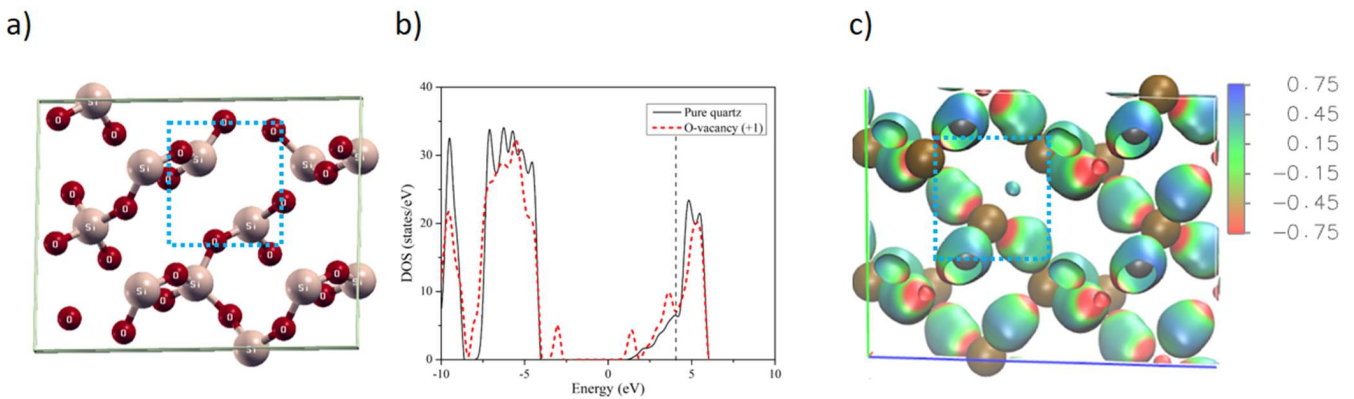

Fig-4 (a) The minimum energy structure of a 1x2x2 supercell for +1 charged ODC. The small grey ball represents the H atom and (b) is the DOS plot for the defect. (c) The electrostatic potential on isodensity (0.008 a.u.) surface.

v) Another possibility is that an O atom takes two electrons from one from each Si atom. This leaves only paired electron with Si. This defect is simulated by giving an overall +2 charge to the supercell. The minimum energy structure is shown in Fig-5(a). The structure seems to be considerably distorted from pure quartz. On close observation, the distortion results due to moving of Si-Si atoms away from each other (from 3.12 Å to 4.45 Å). The Si atoms also form a weak bond with nearby O-atoms (1.82 Å). The DOS for this defect is shown in Fig-5(b). Significant change in the valence and conduction band is observed due to large distortion in the structure. However, no defect states are observed in the DOS suggesting absence of traps or recombination centres due to the absence of unpaired electrons in Si atoms. The electron density (0.04 a.u. isovalue) is displayed in Fig-5(c). The absence of traps suggests less possibility of showing Luminescence. However, it is important to note that such defect is not yet reported in the literature.

a)
b)
c)

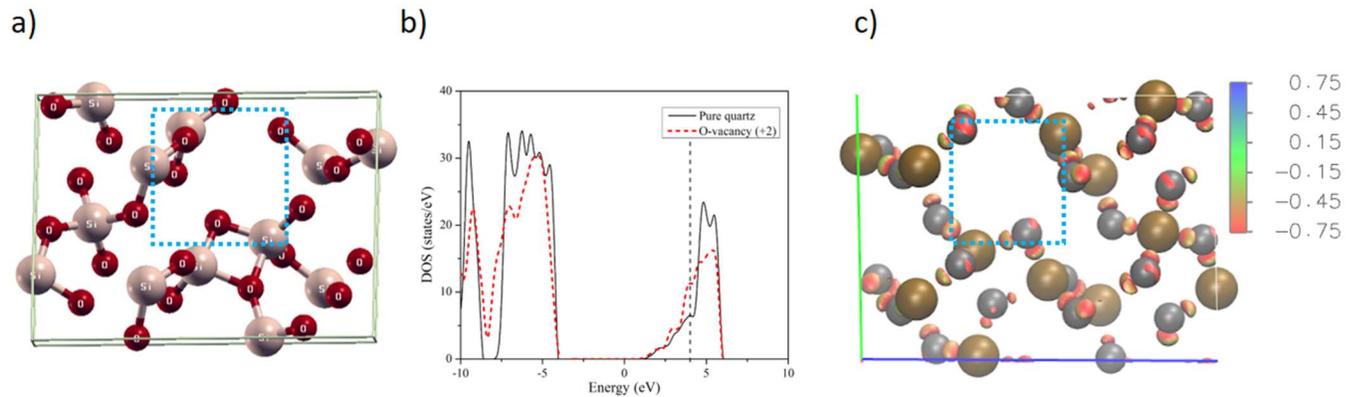

Fig-5 (a) The minimum energy structure of a 1x2x2 supercell for +2 charged ODC. The small grey ball represents the H atom and (b) is the DOS plot for the defect. (c) The electrostatic potential on isodensity (0.04 a.u.) surface.

vi) The weak Si-Si bond formed in case of ODC can be partially or completely passivated due to presence of hydrogen atoms [51]. Partial passivation of the bond is obtained when an H-atom is present near one Si radical (≡SiH-Si≡) formed because of the absence of oxygen. The defect thus created represents the $E_2'$ centre discussed in the introduction and studied extensively by previous researchers [52]. A limited study exists for $E_2'$ centres and the correlation between luminescence and ESR is not yet established. The minimum energy structure for this defect is displayed in Fig-6(a). The distance between the passivated Si atom and H is 1.56 Å with the Si-Si distance increasing to 3.5 Å. The distance between the Si radical and O remains 1.65Å while HSi-O bond length reduces to 1.63 Å. Fig-6(b) displays the DOS for this defect (≡SiH-Si≡). The peak at 2.75 eV in Fig-3(b) disappears and a new peak is observed near the Fermi energy (~-0.1eV) of pure quartz having 5 states/eV DOS. The electron density isosurface (0.04) is shown in Fig-6(c). The electron density between the Si-H bond is the same as the Si-O bond. In addition, the potential between Si-H is slightly positive due to the presence of the H atom. This suggests that it will act as a deep trap and possibly recombination centre, thus, play a role in luminescence emission.

a)
b)
c)

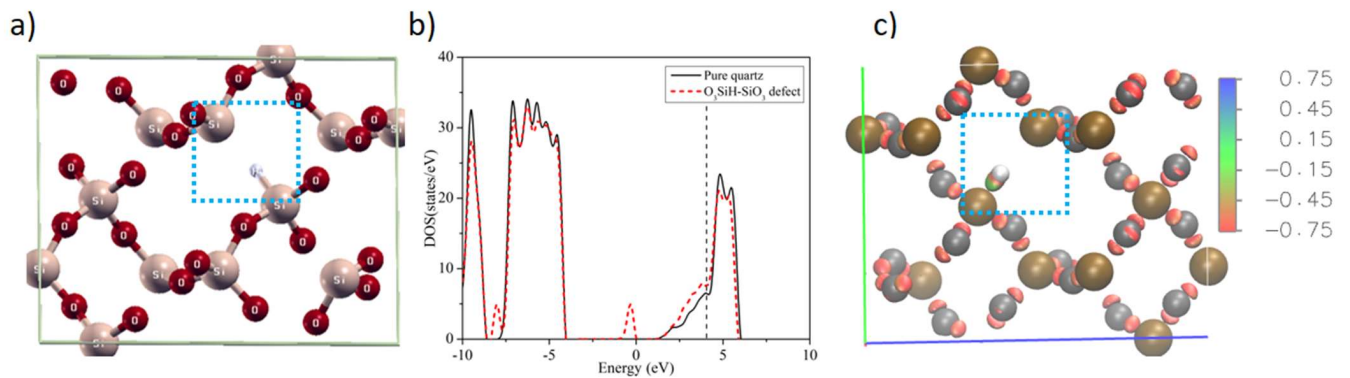

Fig-6 (a) The minimum energy structure of a 1x2x2 supercell for ≡SiH-Si≡ defect. The small grey ball represents the H atom and (b) is the DOS plot for the defect. A peak near the Fermi level is observed. (c) The electrostatic potential on isodensity (0.04 a.u.) surface. Green colour represents positive electrostatic potential.

vii) The structure for fully passivated Si-Si Bond (ODC) is displayed in Fig-7(a). It is constructed by placing two H atoms near the two Si radicals formed in the ODC defect (Fig-3(a)). In this case, the Si-H bond length is 1.45 Å, which is less than that observed in the previous case (≡SiH-Si≡), which indicates that the Si-H bond is stronger in this defect. Si-Si distance is increased to 3.7 Å. The HSi-O bond length changes to 1.62 Å, which is almost the same as the Si-O bond length in pure quartz. The DOS for this defect is plotted in Fig-7(b) which is very similar to that of the pure quartz. An increase in the DOS (~8.8 states/eV) near the shoulder of the conduction band is observed. The electron density isosurface (0.04 a.u.) for the defect shows that the density between the two Si atoms near the defect is symmetric and is slightly electropositive (Fig-7(c)). The density near the H atoms is slightly more than O atoms. The diminishing of any isolated metastable states in the DOS indicates that if this type of defect is present then it will not respond to luminescence (TL or OSL) as there will be no metastable states responsible for trapping of charges. This could be one of the possible reasons for obtaining low luminescence sensitivity in quartz. The DOS of the defect is investigated for first time and is not yet reported anywhere. The reason could be its inactive nature for OSL and ESR as predicted by DOS. The inclusion of this defect neutralizes the crystal. This could possibly be linked to luminescence sensitivity of dull samples.

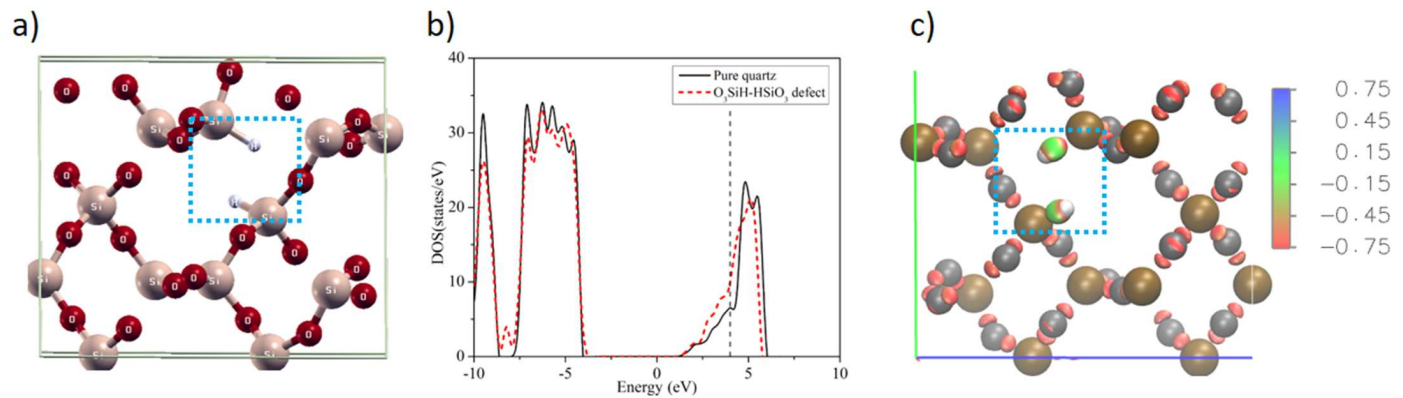

Fig-7 (a) The minimum energy structure of a 1x2x2 supercell for ≡SiH-HSi≡ defect. The small grey ball represents the H atom and (b) is the DOS plot for the defect. (c) The electrostatic potential on isodensity (0.04 a.u.) surface. Green colour represents positive electrostatic potential.

viii) Since in earth's crust, quartz crystallizes in $H_2O$ medium there is a possibility of hydrolysis of the SiO bond resulting in incorporation of OH- point defects in quartz. Thus, OH- may also attach itself to Si radicals formed due to absence of a nearby O atom (ODC). Fig-8(a) shows the optimized structure having one OH- centre attached to one Si leaving the other Si as a radical. Thus, one valence electron of the Si atom does not participate in any bond formation. The distance between two Si atoms (in the proximity of OH-) increases to ~3.6 Å. The bond length of Si-OH is ~ 1.66 Å. The Si-O bond length adjacent to the point defect (the Si atom attached to OH- is 1.61 Å) while that for Si with an unbonded electron is 1.65 Å. The DOS for this defect is displayed in Fig-8(b). Majorly DOS remains unchanged, however a small peak of ~4.6 states/eV observed near the Fermi energy (~0.03 eV above Fermi energy) of pure quartz is formed indicating the formation of metastable state. This DOS is very similar to the one observed for ≡SiH-Si≡ defect and additional states seen near the Fermi energy for both these defects may be because of one unpaired electron on Si atom. In the electron density 0.04 a.u. isosurface plot (Fig-8(c)); although very similar to pure quartz, it is asymmetric near the defect. The electron density between O-H is higher (0.07 a.u.) than Si-O bonds (0.05 a.u.). The presence of radical results in bond structure similar to previous defect E4' centre. This presence of radical results in formation of a deep trap possibly acting as a recombination centre. The speciation of water in quartz is studied extensively using FTIR. Some of the recent finding (Sharma et al 2017) suggest that presence of water in quartz (Si-OH) results in loss of luminescence sensitivity, further the linkages of water presence in quartz suggest the weakening of crystal structure [53-55]. In case of single OH a deep trap is observed, which suggests a presence of slowly decaying OSL signal as observed in some of the quartz separated from rocks. In other cases, no signal is observed in case of quartz and that could be possible due to complete hydrolysis of Si-OH bond.

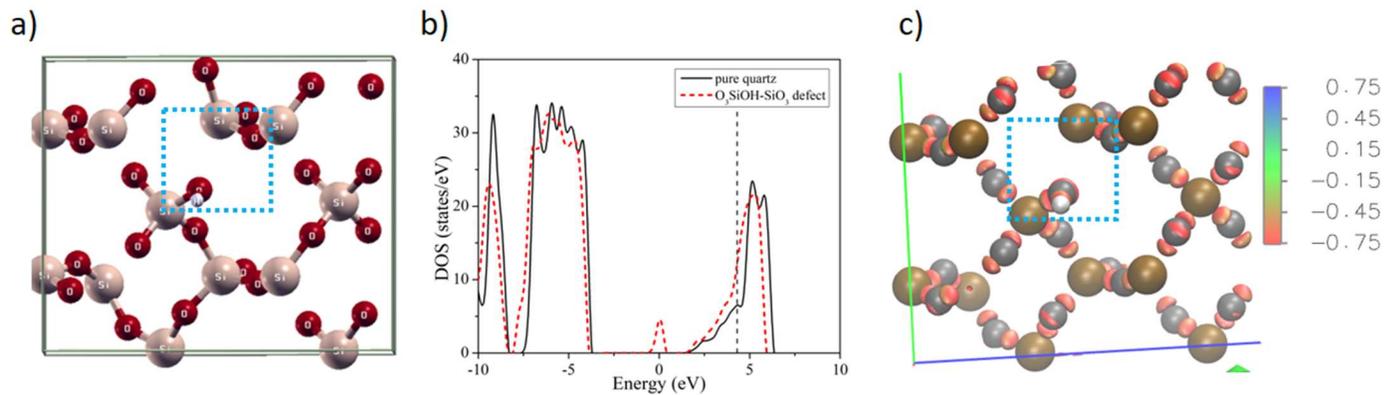

Fig-8 (a) The minimum energy structure of a 1x2x2 supercell for ≡SiOH-Si≡ defect. The small grey ball represents the H atom and (b) is the DOS plot for the defect, which shows a small peak near the Fermi energy of pure quartz. (c) The electrostatic potential on isodensity (0.04 a.u.) surface.

ix) Further hydrolysis of Si-O bond can result in formation of 2 OH- on neighbouring Si atoms resulting in formation of silanol (SiOH- - -HOSi) [56]. The minimum energy structure of such a defect where the Si-Si bond formed due to O vacancy, passivated by OH- is displayed in Fig-9(a). Si-OH bond length is 1.74 Å, which is greater than that observed in the previous case. However, Si-Si distance is reduced to 3.0 Å. Also a very weak H-bond is also formed with the nearby O atom (~2.0 Å). Fig-9(b) shows the plot of DOS for ≡SiOH-HOSi≡ defect. This suggests that the presence of this defect does not alter the band structure of quartz. The electron density isosurface (0.04 a.u.) for the defect is displayed in Fig-9(c). The density surface is seen near the Si-O bonds and O-H bonds. Red colour indicates negative (~0.75 eV) potential. As seen in the previous defect, the O-H bond is stronger than the Si-O bond suggested by larger electron density. This indicates that in such a case there will be no metastable states hence no trapping of charges resulting in no luminescence (TL/OSL) response. As most of the natural rocks are crystallized in presence of water, there is a significantly high probability of existence of SiOH bond in crystallized quartz rocks. It is often observed that the luminescence sensitivity of rocks is quite low. However, the sediment derived from the same rock after natural processing or annealing in the laboratory enhances its luminescence sensitivity by several orders of magnitude. One of the possibilities could be loss of OH defect during processing resulting in enhanced sensitivity. Generally the OSL decay curve of well-behaved quartz decays very fast such that its luminescence is reduced to residual levels in <1 sec. However, in other types of quartz either a slow decaying OSL curve is observed or no luminescence is observed. Present results suggest that slow decaying OSL curve could be the result of presence of single OH bond with a radical resulting in formation of a deep trap while the curve showing no luminescence could be because of presence of silanol group.

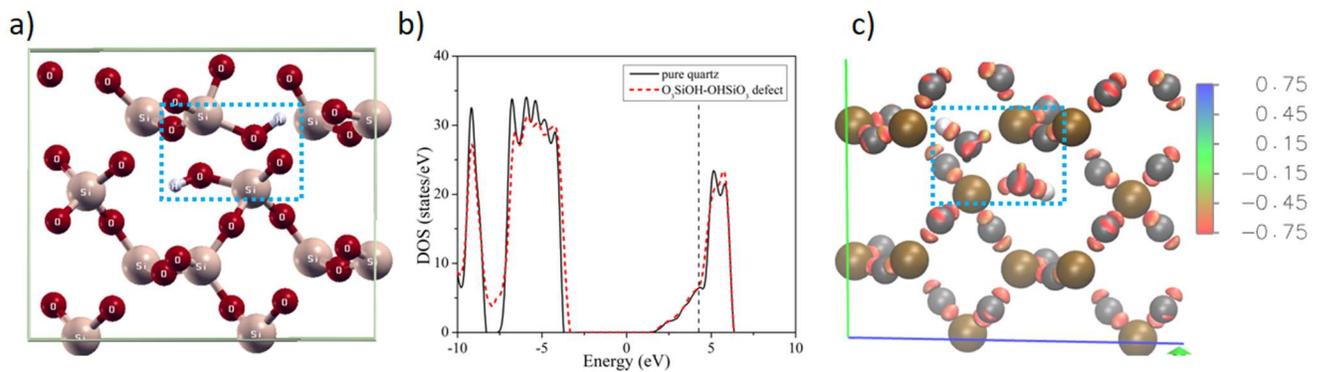

Fig-9 (a) The minimum energy structure of a 1x2x2 supercell for =SiOH-HSi= defect. The small grey ball represents the H atom and (b) is the DOS plot for the defect. (c) The electrostatic potential on isodensity (0.04 a.u.) surface. Green colour represents positive electrostatic potential.

x)  The ODC may create both hole and electron traps [46]. If an electron is trapped near the defect centre, one electron would remain unpaired in the crystal. This creates a -1 overall charge. This defect is studied by giving -1 total charge to the supercell in the input. The minimum energy structure for this defect is displayed in Fig-10(a). In this case, the Si-Si distance increases to 2.65 Å. The Si-O bond length near the defect is ~1.7 Å. The DOS of the defect is displayed in Fig-10(b). In this case, two additional peaks are observed near the valence and conduction band with DOS ~4.8 states/eV. One peak is ~0.8 eV above the valence band maxima and another is ~3.34 eV below the reference (4eV). At 2.5 eV, the DOS considerably increases from 1.9 states/eV to 5.8 states/eV. The electron density near the Si-Si is reduced. As seen in Fig-10(c) the density between Si atoms is ~0.02 a.u. with positive potential at the centre. The newly observed metastable states near the valence band is possibly a hole-trapping centre with low trap depth. However, it is interesting to see formation of new metastable states below the conduction band that can capture electrons and extension of the band tail states. The trap depth of metastable states from reference quartz state is quite high resulting in stable state formation, extension of band tail states may be localized in small area near point defect, thus may not have influence over other parts of the crystal. Nevertheless, such coexistence of metastable state and bandtail states can possibly result in tunnelling phenomena leading to anomalous fading of luminescence signals.

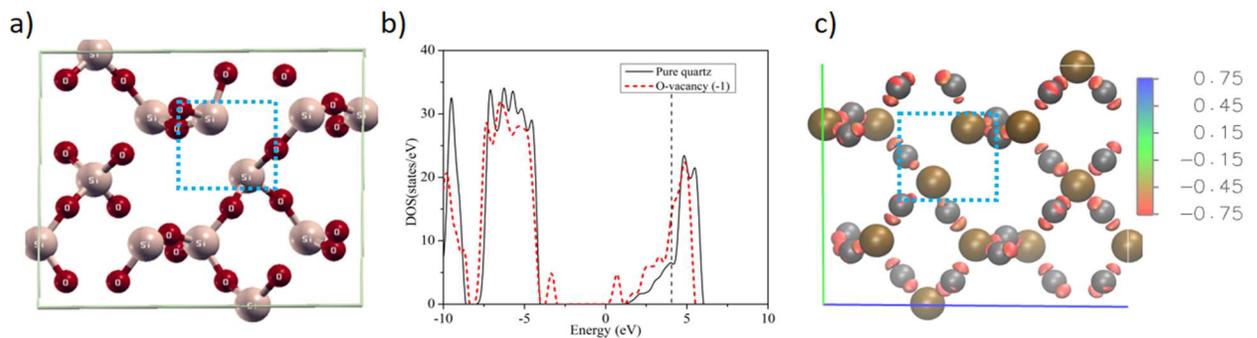

Fig-10 (a) The minimum energy structure of a 1x2x2 supercell for -1 charged ODC. The small grey ball represents the H atom and (b) is the DOS plot for the defect. (c) The electrostatic potential on isodensity (0.04 a.u.) surface.

xi) The minimum energy structure for charge state -2 ODC, i.e. when two electrons are trapped near the ODC, is displayed in Fig-11(a). This defect is not reported in literature but according to DFT, there is a non-zero possibility of formation of such defect, so the defect is simulated for present work. In this charged state, the Si-Si distance increases to 3.6 Å (greater than pure quartz). The Si-O bond length near the defect also increases to 1.77 Å. Fig-11(b) displays the DOS for this defect Both the conduction and valence bands are shifted to lower(~1.52 eV) energies and two peaks (at -2.35 (-0.975) eV and -0.77 (0.60 eV) with 5 states/eV are observed lying below the Fermi energy of pure quartz. Looking at the density plot (Fig-11(c)), it is observed that near the defect the density of electrons is reduced, which indicates that there is no bond formation between the two Si atoms as the additional two electrons will passivate the Si atoms. The density between Si and O near the defect also reduces, suggesting weaker Si-O bonds. This indicates there is deficiency of positive charges so the two metastable states formed will be hole centres. The trap depth of these states indicate one of these will act as a hole trapping centre and another one will act as a recombination centre.

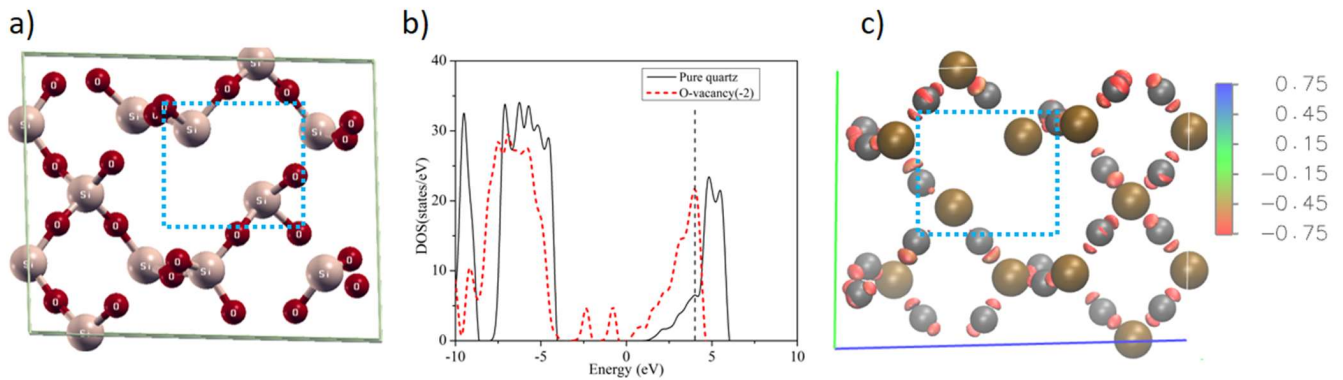

Fig-11 (a) The minimum energy structure of a 1x2x2 supercell for -2 charged ODC. The small grey ball represents the H atom and (b) is the DOS plot for the defect. (c) The electrostatic potential on isodensity (0.04 a.u.) surface.

## (B) Silicon Vacancy Defect:

Si-vacancy is another point defect, which is present in natural quartz during crystallization. As Si is a major element, responsible for sharing 4 electrons in $SiO_2$ structure, the vacancy of Si can significantly affect the crystal field and energy levels surrounding the defect site. Si vacancies are not as common as O-vacancies and interstitial Si has never been reported [40]. There can be several defects, which are related to Si vacancy defects.

i) The Si vacancy point defect is simulated by removing one Si atom in the 1x2x2 supercell. The minimum energy structure for Si vacancy is displayed in Fig-12(a). It is observed that Si vacancy creates two peroxy bridges between Si atoms (Fig. 12a). Capron et al in 2000 studied Si-vacancy in quartz using LDA and GGA and reported formation of two peroxy bridges [57]. SiO-OSi bond length is ~1.48 Å. The distance between the Si-atom and the O belonging to the peroxy bond is 1.69 Å (1.70 Å reported in Capron et al [57]), while the remaining Si-O (remaining 3 O) bond length is slightly increased to 1.63 Å compared to pure quartz. The DOS of this defect structure shows three small peaks near the valence band (Fig-12(b)). The peak closest to the valence band (~0.4eV above the valence band) has ~4.5 states/eV DOS value. The other two peaks (~5 states/eV) are located at 1.8eV and 2.5 eV (above the valence band energy of pure quartz). Looking at the electron density (isovalue ~0.04 a.u.) plot displayed in Fig-12(c), the volume of isosurface increases near the region with Si vacancy (peroxy bonds). Moreover, the electron density near the O atoms of the peroxy bonds is higher (~0.067 a.u.) with a negative electric potential. Thus, both the DOS and electron density analysis indicate the presence of hole traps near this defect centre. Si vacancy hole centres in quartz have not yet been characterized in detail [58]. The deficiency of Si will result in unpaired electrons with Oxygen, which result in peroxy bond formation and creation of hole trapping centres. In some of the recent literature, peroxy pair is shown to be linked with luminescence sensitivity and luminescence/ESR based provenance studies; however, its linkage to luminescence sensitivity and thus to provenance is difficult to establish. It may happen that particular PT crystallization conditions favours creation of such vacancy but its correlation with luminescence sensitivity may not be appropriate and it may be a different proxy altogether not at all linked with luminescence properties. The existence of Si vacancy related recombination centres may however result in enhancement of certain emission bands, which can be linked to peroxy defects.

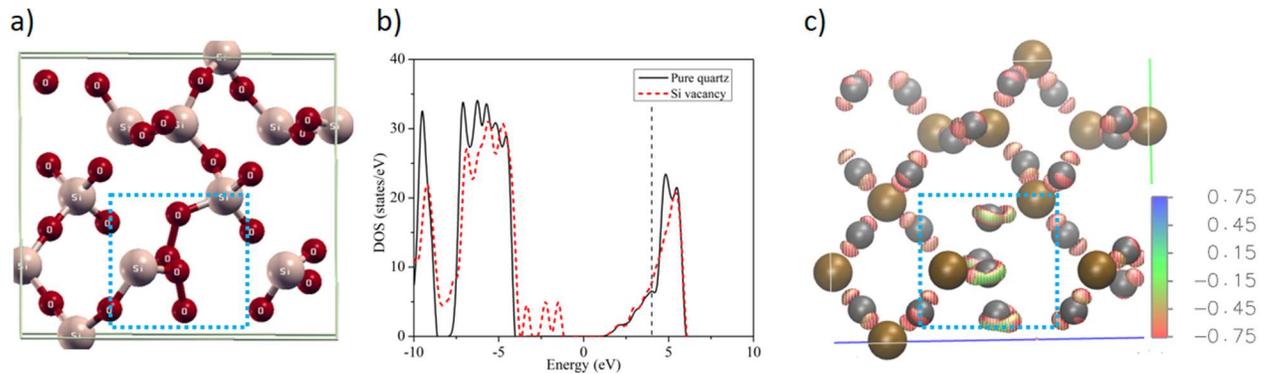

Fig-12 (a) The minimum energy structure of a 1x2x2 supercell in case of Si-vacancy and (b) is the DOS plot for the defect. (c) The electrostatic potential on isodensity (0.04 a.u.) surface.

ii) A silicon vacancy may also host $2H^+$ ions. Thus two of the four dangling O atoms formed due to absence of one Si atom is stabilized. This defect is formed by removing a Si atom and putting 2H atoms close to two unbonded O atoms in the supercell. The stable structure having minimum energy is displayed in Fig-13(a). A peroxy bond between the O atoms having unpaired electrons is observed with a bond length of 1.49 Å. The H+ ions form OH bonds with length 0.98 Å. These H atoms also form a weak hydrogen bond with the nearby O atom. These weak H bonds stretch the Si-O bond length to 1.7 Å. The DOS shows the formation of a metastable states below conduction band which is lying ~4eV below the reference (Fig-13(b)). The electron density plot is similar to Si-vacancy except there is only one peroxy bond (Fig-13(c)). The density of electrons near the O of the proxy bond is comparatively large (~0.1) suggesting it will attract a hole. This looks like a deep electron trap with a trap depth of 4 eV. This may participate in the luminescence recombination process.

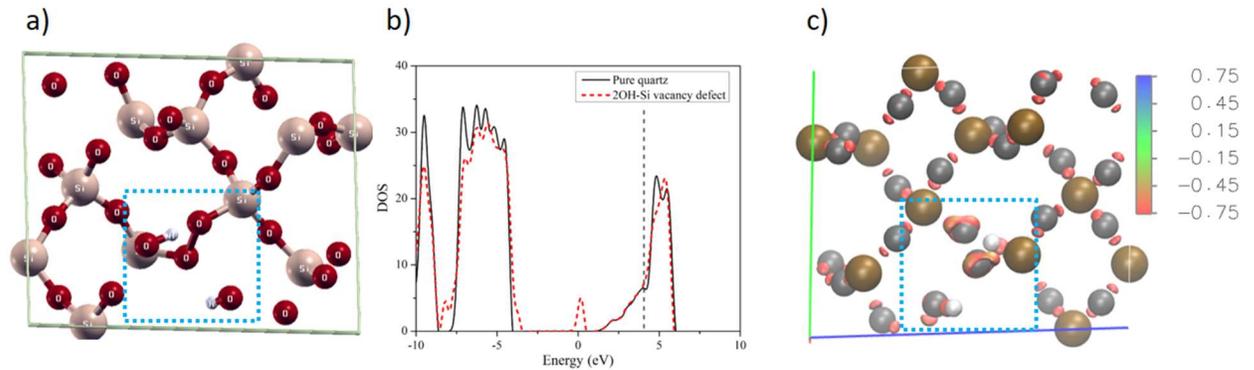

**Fig**-13(a): The minimum energy structure of a 1x2x2 supercell in case of Si-vacancy, (b) is the DOS plot for the defect and (c) is the electrostatic potential on isodensity (0.04 a.u.) surface.

iii) A silicon vacancy can also be substantiated by three or four H ions [40]. Similar to the previous defect, $H_3O_4$ and $H_4O_4$ defects are created by adding three and four atoms, respectively near the unbonded O atoms [40]. The minimum energy structures are displayed in Fig-14(a, d). The Si-O bond length in case of $H_3O_4$ is 1.64 Å, while for the unbonded O; Si-O distance is 1.62 Å. In the case of $H_4O_4$, the Si-O distance is ~1.62 Å that is similar to that of pure quartz. Fig-14(b, e) displayed the DOS for $H_3O_4$ and $H_4O_4$ defect, respectively. The DOS is similar to that of pure quartz in both cases. This suggests that even three H ions may completely passivate the Si vacancy. This result is in contradiction with what is reported [18]. According to [18], a recombination centre (electron trap) is formed near this defect, thus a peak should appear near the Fermi energy in the DOS. Looking at the electron density plot of $H_3O_4$ in Fig-14(c), it is similar to pure quartz. The density near the unpaired O atom is low (0.02) and makes weak H-bonds with the nearby H atoms. Several researchers have associated luminescence emission with $H_3O_4$ centres in blue [59]. Fig 14-(f) displays the density for $H_4O_4$ defect for 0.04 isovalue, which is also similar to the pure quartz except some redistribution due to the presence of 4 H atoms instead of Si. This is interesting, as some of the previous research works as discussed in the introduction associate UV and blue emission in quartz with $H_3O_4$ defect. In quartz it was concluded using electron spin resonance that the 380 nm emission was due to the recombination of electrons with holes trapped at H3O4 centres [12]. However, The ESR studies by Yang and Mckeever [59] and [60] also reported the presence of this centre at 25K. In this defect, the g value is observed at 2.0012, which could be possible due to the presence of one unpaired electron of the O radical. However, DFT simulations suggest that such an arrangement removes the metastable states, thus resulting in no emission at all. Normally the luminescence studies are carried above room temperature at which $H_3O_4$ defects are not expected to be observed and hence cannot be associated with the UV/blue emission observed in luminescence emission. This needs to be understood further.

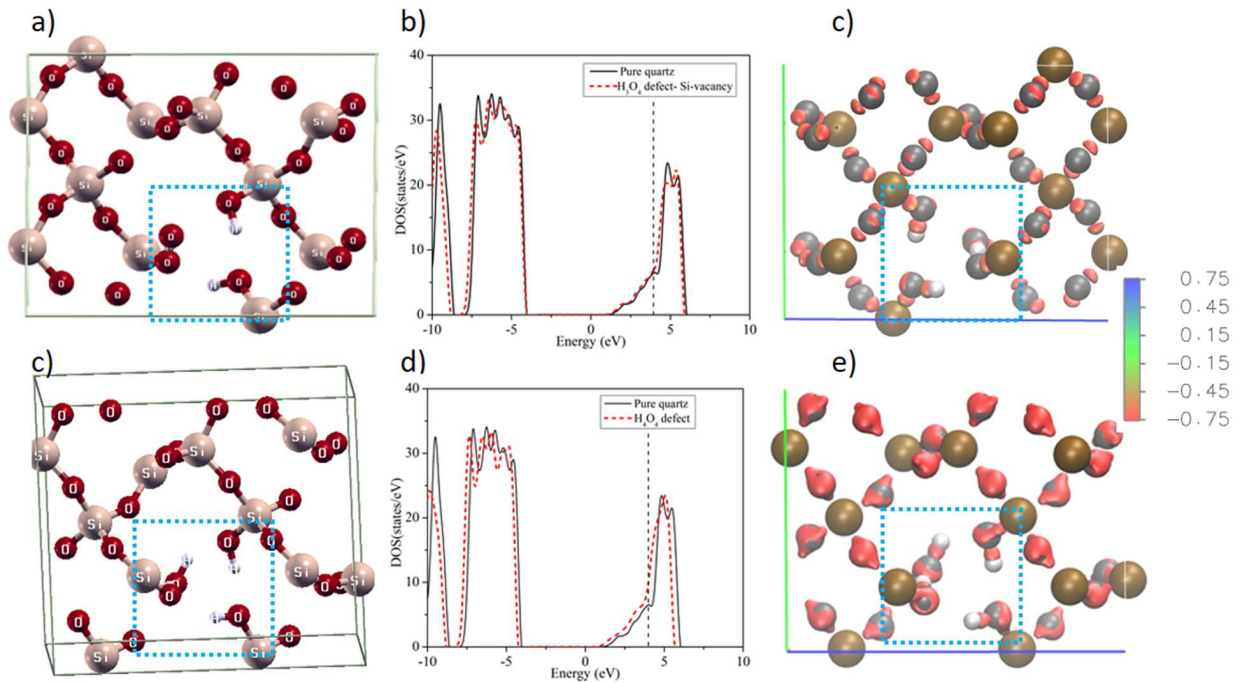

Fig-17 (a) The minimum energy structure of a 1x2x2 supercell in case of $H_3O_4$ defect and (b) is the DOS plot for the defect. (c) The electrostatic potential on isodensity (0.04 a.u.) surface for $H_3O_4$ defect. (d) The minimum energy structure of a 1x2x2 supercell in case of $H_4O_4$ defect and (e) is the DOS plot for the defect. (f) The electrostatic potential on isodensity (0.04 a.u.) surface for $H_4O_4$ defect

## (C) Oxygen excess defect

Oxygen excess in the quartz crystal also leads to intrinsic defects. Peroxy intrinsic defect centres (≡Si–O–O·) and nonbridging oxygen (NBOHC) are two paramagnetic defects formed due to presence of excess O-atoms which are detected by electron spin resonance [61]. O-excess may also form peroxy linkage (≡Si–O–O–Si≡) in the crystal.

i) We studied the ≡Si–O–O–Si≡ defect in quartz, since this defect is also observed in case of Si vacancy. This defect is introduced in the crystal by replacing one O in the supercell by a peroxy link (O-O). The minimum energy structure for this defect is shown in Fig-15(a). The O-O bond is of length ~1.49 Å (similar to the one observed in case of Si-vacancy). The distance between Si atom and O atom in the peroxy bond is 1.67 Å. Fig-15(b) shows the DOS for ≡Si–O–O–Si≡ defect. Two small peaks, one near the valence band and other near the conduction band, with the same DOS (~5 states/eV) are observed. New states seen near the valence band are ~1eV above the valence band maxima. The other peak (near the conduction band) is observed at energy ~2.86 eV below the reference line. Fig-15(c) displays the electron density isosurface (~0.07). A high electron density (~0.07 a.u.) is seen near the peroxy bond. In addition, some regions near the O atoms of peroxy bonds have negative charge density (-0.05 a.u.) (Figure 15 c, d highlighted box). The oxygen excess peroxy defect is reported in reference to electron spin resonance linked with provenance studies. However, studies related to luminescence are limited. It is however known that presence of excess oxygen gives rise to orange/red emission in luminescence. The electron trap at 2.86 represents presence of deep traps. With the excess oxygen defects, the electron and the hole traps that are created are well separated in energy domain and are formed near to conduction and valance band respectively, but are spatially very near to each other. This signifies high probability of quantum mechanical tunnelling between them leading to a thermal fading of stored charges over time. Generally fading is not reported in quartz, except for volcanic quartz, where red emission (related with Iron or NBOHC or oxygen excess defects) is significant.

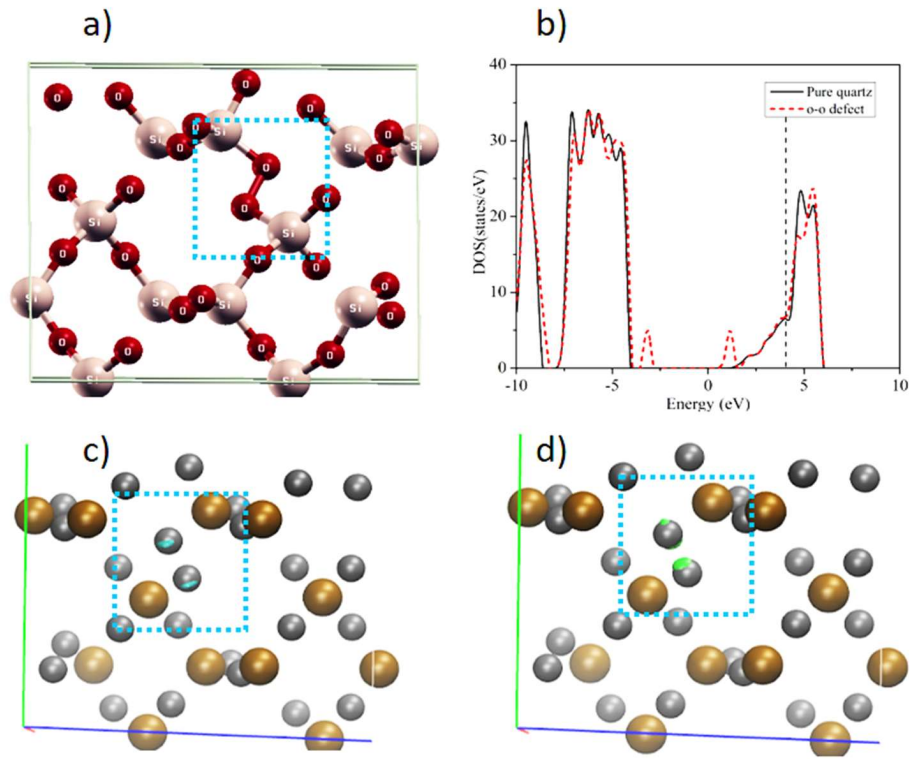

Fig-15 (a) The minimum energy structure of a 1x2x2 supercell in case of ≡Si−O−O−Si≡ defect and (b) is the DOS plot for the defect. (c) The electrostatic potential on isodensity (0.07 a.u.) surface (blue colour) and at (d) -0.05 a.u. (green colour).

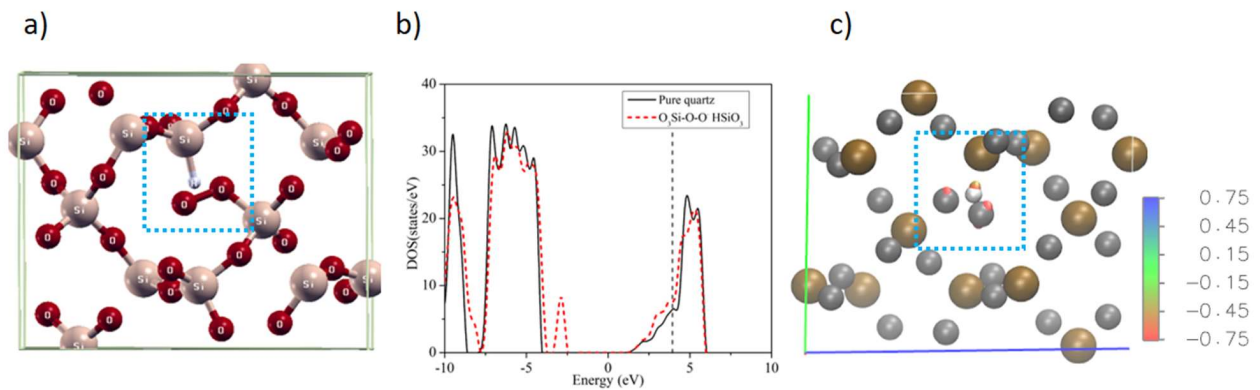

Fig-16 (a) The minimum energy structure of a 1x2x2 supercell in case of ≡Si−O−O· HSi≡ defect and (b) is the DOS plot for the defect. (c) The electrostatic potential on isodensity (0.09 a.u.) surface.

ii)  Peroxy intrinsic defect centres (≡Si−O−O·) is formed by attaching one H atom to a Si atom, leaving one O of peroxy bond dangling [61]. Thus, O atom has one unpaired electron that does not take part in any bond formation. Fig-16(a) displays the optimized structure of this defect. The peroxy bond length is 1.35 Å, which is shorter than ≡Si−O−O−Si≡ defect. The Si-H bond length is 1.46 Å. The Si-O bond length, where one O forms a peroxy bond, is increased to 1.72 Å (1.62 Å for pure crystal). Fig-16(b) shows the DOS for (≡Si−O−O·). A sharp peak is seen near the valence band. This peak is observed 1.12 eV above the valence band maxima of pure quartz with 8.3 states/eV. Some change in the conduction band bottom is also seen. Fig-16(c) shows the electron density plot for this defect at isovalue 0.07 a.u. The electron density near O-O· and Si-H is more and the pseudo red colour in the plot indicates it to be electronegative. This suggests formation of a hole trap near to the valence band with higher number of trap availability. The peroxy defect may be linked to 110° C considering its experimentally estimated trap depth (0.98) which is near to DFT results. The observed DOS for the peak is high, and the 110° C peak is often a very intense peak. Further peroxy defects are also observed using the ESR method (ref). This suggests the possibility that 110°C TL peak can be linked to the above-mentioned defect centre.

### (D) Al defect:

As discussed in the Section 1 (Introduction) $Al^{3+}$ is the most common impurity in quartz. We study this defect by replacing a Si atom with an Al atom. The minimum energy structure of the defect is shown in Fig-17(a). The Al-O bond length is ~1.74 Å, greater than Si-O in quartz. The O atoms near the impurity moves towards the nearby Si atom reducing the Si-O bond length to ~1.60 Å. The DOS shown in Fig-17(b) shows no change on the substitution of Si with Al. In the electron density plot (0.04 isovalue), the density between Al-O is slightly reduced (Fig-17(c)) which suggests that Al-O bonds are weaker than Si-O bonds. Since the valency of Al is +3, the charge can be compensated by an addition of a monovalent ion. Here, H+ ion is added to compensate the charge ($[AlO_4]^-[H]^+$). Fig-17(d) shows the minimum energy structure for this defect. In this case, the Al-O bond length is ~1.725 Å while the Al-O-H bond length is 1.90 Å. The DOS in this case is again very similar to that of pure quartz except for some tail states in the valence band. The electron density plot (Fig-17(f)) shows 0.008 a.u. density between Al and O atoms. The DFT results suggest that Al point defects will not alter the electronic structure of quartz and no recombination or trap centres should be observed. However, as discussed in the introduction, electron-hole recombination at $[AlO_4]^0$ or at $[AlO_4/M^+]^+$ centre is believed to emit a UV-violet band [16]. Therefore, it is not clear from the DFT perspective whether suitable states for recombination are available or not? The question is how the recombination and luminescence emission is taking place. This possibly suggests that more refinements are needed in our understanding of luminescence mechanism.

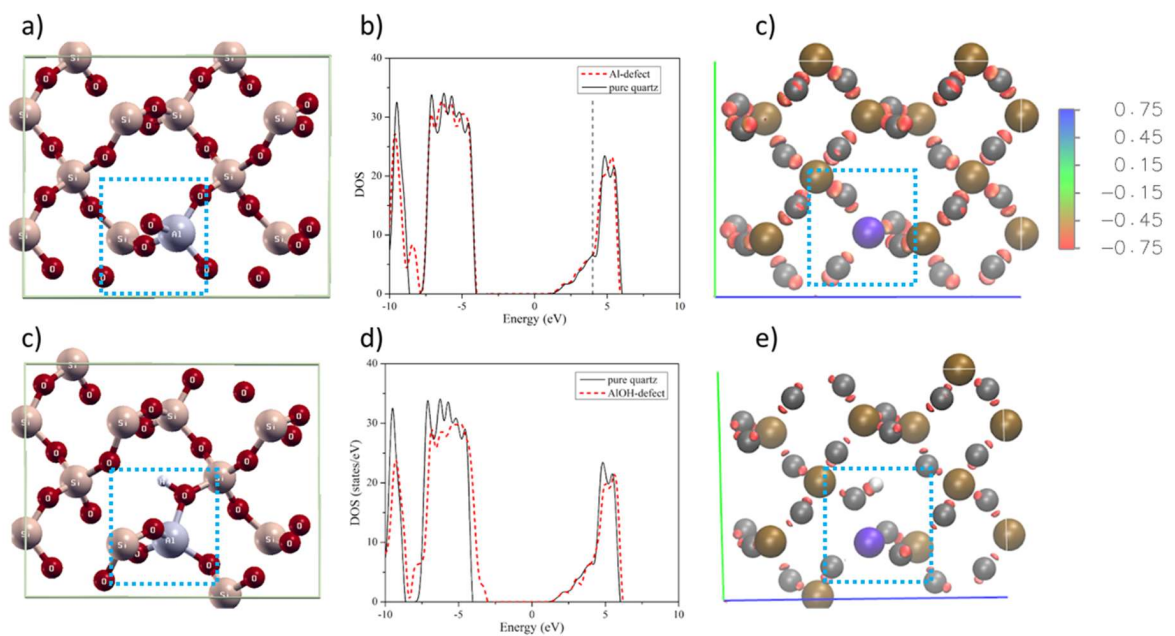

Fig-17 (a) The minimum energy structure of a 1x2x2 supercell in case of [AlO4]⁻ defect (Al atom is represented by grey ball) and (b) is the DOS plot for the defect. (c) The electrostatic potential on isodensity (0.04 a.u.) surface for [AlO4]⁻ defect. (d) The minimum energy structure of a 1x2x2 supercell in case of [AlO₄]⁻[H]⁺ defect and (e) is the DOS plot for the defect. (f) The electrostatic potential on isodensity (0.04) surface for [AlO₄]⁻[H]⁺ defect

### (E) Fe-defect:

Fe being the most abundant element on earth's crust is a commonly found defect in pure quartz. This defect is studied by substituting Si by Fe. However, due to larger size, incorporation of Fe ion in lattice produces significant strain in the crystal lattice and its incorporation is difficult as compared to Si or Al ions. Fig-18(a) shows the optimized structure of this defect. The Fe-O bond length is ~1.74 Å (same as Al-O), greater than Si-O in quartz. The Si-O (near Fe atom) bond length is ~1.65 Å which is more than that for pure quartz (~1.61 Å). The DOS for Fe-defect is shown in Fig-18(b). Two large peaks are observed below the Fermi energy of pure quartz at -2.56 eV (9.5 states/eV) and -1.61 eV (12 states/eV). The electron density plot (isovalue 0.04 a.u.) is displayed in Fig-16(c). From the DOS it is inferred that these are metastable states for holes. Fe is often linked with red emission especially in the volcanic quartz [1]. The deep traps at -2.56eV may act as recombination centres, while the shallower traps can act as hole centres suitable for dosimetry.

In this case, as well, the Fe charge can be compensated by adding an ion. The structure for this defect [FeO₄]⁻H⁺ is shown in Fig-18(d). Here, the Fe-O bond length is ~1.84 Å while the Fe-O-H bond length is 1.92 Å. A large band below the Fermi energy of pure quartz is observed in the DOS of [FeO₄]⁻H⁺ (Fig-18(e)) with two peaks (-1.9 eV and -0.55 eV). The electron density plot for [FeO₄]⁻H⁺ (isovalue 0.04 a.u.) is displayed in Fig-16(f). The introduction of H⁺ further modifies the trap recombination centres. In this case, both the observed traps will act as hole trapping centres.

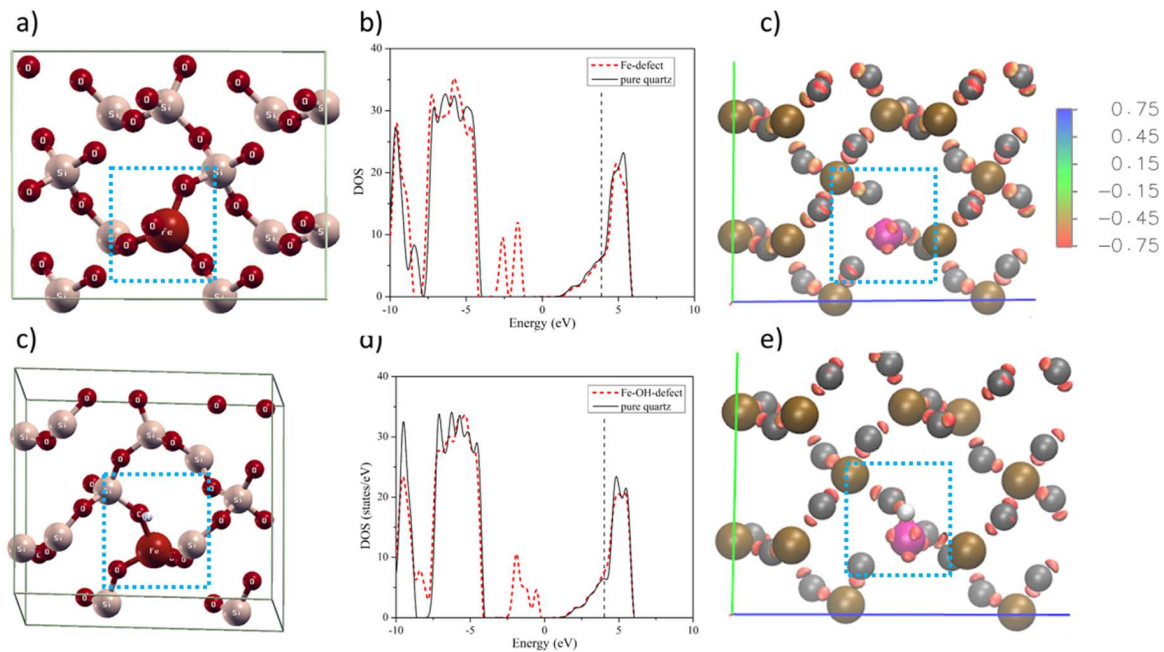

Fig-18 (a) The minimum energy structure of a 1x2x2 supercell in case of [FeO4]⁻ defect (Fe atom is represented by an orange ball) and (b) is the DOS plot for the defect. (c) The electrostatic potential on isodensity (0.04 a.u.) surface for [AlO4]⁻ defect. (d) The minimum energy structure of a 1x2x2 supercell in case of [FeO₄]⁻[H]⁺ defect and (e) is the DOS plot for the defect. (f) The electrostatic potential on isodensity (0.04) surface for [FeO₄]⁻[H]⁺ defect

## V. Summary and Conclusions

This study explores the effect of addition of defects on quartz crystal with the help of DFT. The DOS and electron density is studied for most commonly found intrinsic defects in quartz crystal. The major findings of the study are summarized in Table 2. From the DOS and electron density analysis, we predict which particular intrinsic defect is responsible for the formation hole and electron trapping and recombination centres. It is observed that the intrinsic defects play a crucial role in governing the luminescence properties of the quartz crystal. Following observations are seen.

1. Multiple electron traps ranging from a trap depth of 1.25- 4eV are formed.
2. Multiple hole traps ranging from 0.03 eV to 2.25eV are formed.
3. The addition of more H or OH in the crystal structure reduce the traps and hence could be responsible for reduction in luminescence sensitivity.
4. Generally, defects either create only an electron trap or a hole trap; however, Silicon vacancy (leading to peroxy bonds) or an excess oxygen defect creates both the electron and hole defects.
5. With the excess oxygen defects, the electron and the hole traps that are created are well separated in energy domain and are formed near to conduction and valence band respectively, but are spatially very near to each other.
6. It is reported that UV and blue emissions are due to $H_3O_4$ defects, but in DFT simulations no traps/metastable is observed in the DOS for the presence of these defects.
7. For extrinsic defect such as Al, negligible change in DOS is observed, whereas Fe majorly creates recombination and hole centres.

Table 2: Summary of DFT finding of the simulated defects and corresponding luminescence inference.

| Defect possibility | Possible defect combination | DFT finding | Inference from luminescence perspective |
|---|---|---|---|
| Oxygen vacancy | Oxygen missing & Si-Si bond | A peak is formed, 1.25 eV below our reference conduction band bottom | Neutral oxygen vacancy may acts as electron trap. The reported lifetime of a trap having trap depth ~1.42 eV with TL peak at 190 °C [7] is around 700 years suggesting the observed trap with 1.25 eV trap depth will have a lower lifetime. The presence of low-density band tail below CB states may also result in fading of the signal due to leaking of charges for overtime depending upon ambient temperature reducing the stability further. |
|  | Oxygen missing and an electron from Silicon missing | In the DOS (Fig-4(b)) a new peak appears at ~0.3 eV below the reference. Two other small peaks are seen, one at ~1eV | The peak near VB is found to be matching with the reported activation energy (~0.95eV with lifetime $10^8$ years) [41] . However, there is no |

| | | above the valence band maxima and other at ~2.6 eV below the reference level, both having ~5 states/eV DOS. | report about 2.6eV trap centre. |
|---|---|---|---|
| | Oxygen missing and an electron from each of the two Silicon missing | Significant change in the valence and conduction band is observed due to large distortion in the structure. However, no defect states are observed in the DOS suggesting absence of traps or recombination centres due to the absence of unpaired electrons in Si atoms. | The absence of traps suggests less possibility of showing Luminescence. |
| | Oxygen missing & H attached to one silicon | a peak is ~4eV below the bottom of the conduction band | DFT suggests that it will act as recombination centre. However, it is widely used in EPR dating, known by the name E' centre. |
| | Oxygen missing & an H attached to each of the two silicon | No distinct new peak is observed; An increase in the DOS (~8.8states/eV) near the conduction band is observed. | The addition of more H in the crystal structure, diminishes the traps and hence will act to reduce the luminescence sensitivity |
| | Oxygen missing & OH attached to Si | a small peak of ~4.6 states/eV observed near the Fermi energy (~0.03 eV above Fermi energy) of pure quartz | In case of single OH a deep trap is observed, which suggests a presence of slowly decaying OSL signal as observed in some of the quartz separated from rocks. In other cases, no signal is observed in case of quartz and that could be possible due to complete hydrolysis of Si-OH bond. |
| | Oxygen missing & 2OH attached to Si | No change in the DOS. Fig 9 | Addition of more OH defect diminishes the traps and quenches luminescence properties. |
| Silicon vacancy | Silicon missing | The DOS of this defect structure shows three small peaks near the valence band (Fig-12(b)). The peak closest to the valence band (~0.4eV above the valence band) has ~4.5 states/eV DOS value. The other two peaks (~5 states/eV) are located at 1.8eV and 2.5 eV (above the valence band energy of pure quartz). | Silicon vacancy acts as hole trap |

| | Silicon missing: 4H/3H replacing Si | Fig-14(b, e) displayed the DOS for H 3 O 4 and H 4 O 4 defect, respectively. The DOS is similar to that of pure quartz in both cases. | According to [18], a recombination centre (electron trap) is formed near this defect, thus a peak should appear near the Fermi energy in the DOS. However, DFT predicts that this defect should not result in any trapping center. |
|---|---|---|---|
| | Silicon missing: 2H replacing Si and 1 peroxy bond | The DOS shows the formation of a metastable states below conduction band which is lying ~4eV below the reference (Fig-13(b)). | This looks like a deep electron trap with a trap depth of 4 eV. This may participate in the luminescence recombination process. |
| Substitutional defect | Silicon replaced by Aluminium +1H/Li/Na | The DOS in this case is again very similar to that of pure quartz except for some tail states in the valence band. | Should not be related with luminescence characteristics, however many studies have shown that Al defect creates metastable states [41] |
| | Silicon replaced by Aluminium | | |
| | Silicon replaced by Iron (Iron present in 3+ oxidation state) | Two large peaks are observed below the Fermi energy of pure quartz at -2.56 eV (9.5 states/eV) and -1.61 eV (12 states/eV). | In this case, both the observed traps will act as hole trapping centres. |
| Excess Oxygen | Peroxy linkage [$O_3SiOO^-$] | Two small peaks, one near the valence band and other near the conduction band, with the same DOS (~5 states/eV) are observed. New states seen near the valence band are ~1eV above the valence band maxima. The other peak (near the conduction band) is observed at energy ~2.86 eV below the reference line. | The electron trap at 2.86 represents presence of deep traps. With the excess oxygen defects, the electron and the hole traps that are created are well separated in energy domain and are formed near to conduction and valance band respectively, but are spatially very near to each other. This signifies high probability of quantum mechanical tunnelling between them leading to a thermal fading of stored charges over time. Generally fading is not reported in quartz, except for volcanic quartz, where red emission (related with Iron or NBOHC or oxygen excess defects) is significant. |